\documentclass{jps-cp}
\usepackage{txfonts} 
\usepackage{gensymb}
\usepackage{epsfig,graphicx,pst-all}
\title{Three-body Description of 2$n$-Halo and Unbound 2$n$-Systems: $^{22}$C and $^{26}$O}

\author{Jagjit \textsc{Singh}$^{1,2}$, W. \textsc{Horiuchi}$^{3}$, L. \textsc{Fortunato}$^{4,5}$ and A. \textsc{Vitturi}$^{4,5}$}

\inst{$^{1}$Nuclear Reaction Data Centre, Faculty of Science, Hokkaido University, Sapporo 060-0810, Japan \\
$^{2}$Research Center for Nuclear Physics (RCNP), Osaka University, Ibaraki 567-0047, Japan\\
$^{3}$Department of Physics, Hokkaido University, Sapporo, 060-0810 Japan\\
$^{4}$Dipartimento di Fisica e Astronomia \textquotedblleft G.Galilei\textquotedblright, via Marzolo 8, I-35131 Padova, Italy\\
$^{5}$INFN - Sezione di Padova, via Marzolo 8, I-35131 Padova, Italy}
\email{jsingh@rcnp.osaka-u.ac.jp}
\recdate{July 20, 2019}
\abst{We study the two-neutron correlations in the ground state of the weakly-bound two-neutron halo
nucleus $^{22}$C sitting at the edge of the neutron-drip line and  also in the unbound nucleus $^{26}$O sitting beyond the 
neutron dripline. For the present study, we employ a three-body (${\rm core}+n+n$) structure model designed
for describing the two-neutron halo system by explicit inclusion of unbound continuum states of the subsystem (${\rm core}+n$).
We use either a density-independent or a density-dependent contact-delta interaction to describe the neutron-neutron interaction and its strength is 
varied to fix the binding energy. We report the configuration mixing in the ground state of these systems for different choices of pairing interactions.}
\kword{Two-neutron halo, Two-neutron unbound system, Two-neutron correlations.}
\begin{document}
\maketitle
\section{Introduction}
The new generation of radioactive ion beam facilities around the various parts of the globe has 
provided the access to the neutron-rich side of the nuclear chart. Due to this, there has been a rapidly increasing interest in 
the physics of the two-neutron (2$n$) halo nuclei sitting right on the top of neutron driplines and decays of 2$n$-unbound systems beyond the 
neutron dripline. These systems demand a three-body (3b) description with proper treatment of continuum, 
the conventional shell-model assumptions being insufficient. The well established 2$n$-halo $^6$He has long history of studies in 3b-framework 
and thus it can be safely remarked that the understanding of 3b-dynamics is fairly established for $p$-shell nuclei, 
whereas for the $s$-$d$ shell nuclei still the situation is in progress stage \cite{ERSH08}.
In the present study we consider two different $s$-$d$ shell nuclei, the 2$n$-halo $^{22}$C and the 2$n$-unbound system $^{26}$O. 
Very recently a high precision measurement of the interaction cross-section for $^{22}$C 
was made on a carbon target at $235$ MeV/nucleon \cite{TOG} and also the unbound nucleus $^{26}$O has been 
investigated, using invariant-mass spectroscopy \cite{KON} at RIKEN. 
The structural spectroscopy of the two-body subsystem plays a vital role in the understanding the 3b-system. 
These high precision measurements and the sensitivity of the structural spectroscopy of subsystem with the structure of 3b-system (core$+n+n$),
are the motivation for selecting these nuclei for the present study.
We have studied the pairing collectivity in the ground state of the 2$n$-halo $^{22}$C and in the the 2$n$-unbound system $^{26}$O.
For this study we have used our recently implemented 3b-structure model (core$+n+n$) for the ground and continuum states of the 
2$n$-halo nuclei \cite{FOR,JS1,JS2}. We have explored the role of different pairing interactions such as 
density independent (DI) contact-delta pairing interaction and density dependent (DD) contact-delta pairing interaction with the configuration mixing in the 
ground-state of these systems.
\section{Model Formulation}
\vspace{-0.4cm}
In our approach we consider a 3b-system consisting of an inert core nucleus and two valence neutrons, which is specified by Hamiltonian
\vspace{-0.4cm}
\begin{equation}
H=-\frac{\hbar^2}{2\mu}\sum_{i=1}^{2}\nabla_i^2+\sum_{i=1}^{2}V_{{\rm core}+n}(\vec r_i)+V_{12}(\vec r_1,\vec r_2)
\vspace{-0.3cm}
\end{equation}
where $\mu=A_cm_N/(A_c+1)$ is the reduced mass, and $m_N$ and $A_c$ are the nucleon mass and mass 
number of the core nucleus, respectively. The recoil term is neglected in the present study, as $A_c=20$ and $24$ are large enough to ignore it.
$V_{{\rm core}+n}$ is the core-$n$ potential and $V_{12}$ is $n$-$n$ potential. The neutron single-particle unbound $s$-, $p$-, $d$- and $f$-wave continuum 
states of the subsystem $^{21}$C and and $^{25}$O are calculated in a simple shell model picture for the converged model parameter, bin width ($\Delta{E}=0.1$), 
by using the Dirac delta normalization and are checked with a more refined phase-shift analysis. 
These ${\rm core}+n$ continuum wave functions are used to construct the 
two-particle states of the ${\rm core}+n+n$ system by proper angular momentum
couplings. We use a density-independent (DI) contact-delta pairing interaction for simplicity, 
and its strength is the parameter which will be fixed to reproduce the ground-state energy. 
We have also used a density-dependent (DD) contact-delta pairing interaction. For a detailed formulation 
one can refer to \cite{FOR,JS1,JS2}.
\section{Two-body unbound subsystems (${\rm core}+n$)}
\vspace{-0.4cm}
The investigation of the two-body (${\rm core}+n$) subsystem is crucial in understanding the 
three-body system (${\rm core}+n+n$). The interaction of the core with the valence neutron ($n$) 
plays a vital role in the binding mechanism of the ${\rm core}+n+n$ system.
The elementary concern over the choice of a ${\rm core}+n$ 
potential is the scarce experimental information about the core-neutron systems. 
We employ the following ${\rm core}+n$ potential
\vspace{-0.4cm}
\begin{equation}
V_{{\rm core}+n} = \left(V_0^l+V_{ls}\vec{l}\cdot\vec{s}\frac{1}{r}\frac{d}{dr}\right) \frac{1}{1+{\rm\,exp}\left(\frac{r-R_c}{a}\right)},
\label{vsp}
\vspace{-0.4cm}
\end{equation}
\vspace{-0.1cm}
where $R_c=r_0A_c^{\frac{1}{3}}$ with $r_0$ and $a$ are the radius and diffuseness parameter of the Woods-Saxon potential.
\vspace{-0.3cm}
\subsection{$^{21}$C}
\vspace{-0.1cm}
For $^{21}$C, not much is known beyond that it is unbound. 
The only available experimental study using the single-proton removal reaction reported the limit to the scattering 
length $\lvert a_0\lvert < 2.8$\,fm and due to the low statistics of this experimental data at low energies, the possibility of low-lying 
resonance states can not be ruled out \cite{MOSB13}. In the view of exploring the sensitivity of the ${\rm core}$-$n$ potential to the possible resonances and configuration 
mixing in the ground state of $^{22}$C, very recently we examined in detail the four different potential sets (for details see text and Table~1 of Ref.~\cite{JS2}). 
Here we will discuss the results corresponding to potential Set-$1$, which is adopted from the literature \cite{HOR06}, 
beacuse of its acceptance by different $3$-body models leading to good explanation of the observed properties for $^{22}$C.
The subshell closure of the neutron number $14$ is assumed for the core configuration given by $(0s_{1/2})^2(0p_{3/2})^4(0p_{1/2})^2(0d_{5/2})^6$. 
The seven valence neutron continuum orbits, i.e., $s_{1/2}$, $d_{3/2}$, $f_{7/2}$, $p_{3/2}$, $f_{5/2}$, $p_{1/2}$ and $d_{5/2}$
are considered in the present calculations for $^{21}$C.
\vspace{-0.3cm}
\subsection{$^{25}$O}
\vspace{-0.1cm}
In the recent measurement conducted at RIKEN \cite{KON}, along with high accuracy measurement of ground state of $^{26}$O, they have also 
reported the $d_{3/2}$ resonance state at $749(10)$ keV  with width of $88(6)$ keV for $^{25}$O. This information will serve as input for fixing the 
core$+n$ potential parameters. For $^{25}$O, we adopt the same value for the diffuseness parameter ($a$) and the radius parameter ($r_{0}$), as in 
Ref.~\cite{HAG}. For the Wood-Saxon depth parameter ($V_{0}$) and the strength of spin-orbit potential ($V_{ls}$) parameter tabulated in Table~\ref{T2}, 
we use the information for the energy of unbound 
$d_{3/2}$ state. Our parameters for $V_{0}$ and $V_{ls}$ are consistent with the one reported in Ref.~\cite{HAG}. 
The neutron number $16$ is assumed for the core configuration given by $(0s_{1/2})^2(0p_{3/2})^4(0p_{1/2})^2(0d_{5/2})^6(1s_{1/2})^2$. 
The three valence neutron continuum orbits, i.e., $d_{3/2}$, $p_{3/2}$ and $f_{7/2}$
are considered in the present calculations for $^{25}$O.
\begin{table}[]
\vspace{-0.7cm}
\centering
\caption{Parameter sets of the ${\rm core}$-$n$ potential for $l= 1, 2, 3$ states of a $^{24}$O$+n$ system. 
The possible resonances with resonance energy $E_{R}$ and decay width $\Gamma$ in 
MeV are also tabulated.}
\label{T2}
\begin{tabular}{ccccccc}
\hline
$lj$      & $r_{0}$(fm) & $a$(fm) & $V_{0}$(MeV) & $V_{ls}$(MeV)& $E_R$(MeV)&$\Gamma$(MeV) \\ \hline
$d_{3/2}$ &         &        & -44.10        & 22.84        & 0.740 & 0.086  \\
$p_{3/2}$ & 1.25        & 0.72        & -48.67        & 22.84        & 0.570& 1.382   \\ 
$f_{7/2}$ &         &        & -44.10        & 22.84        & 2.440& 0.206   \\ \hline
\end{tabular}
\end{table}
\section{Results and Discussions}
\begin{figure}[]
\vspace*{-0.35cm}
\includegraphics{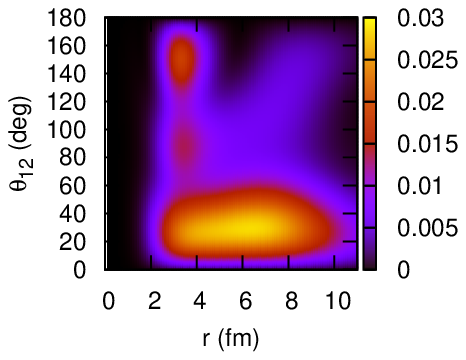}
\includegraphics{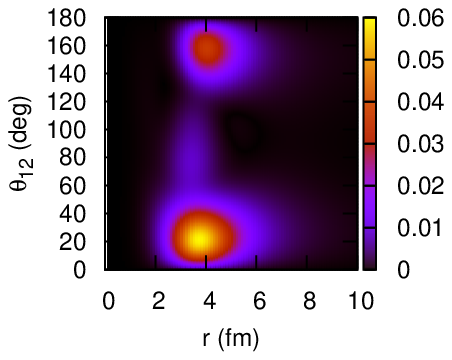}
\caption{Two-particle density for the ground state of $^{22}$C (left-panel) and $^{26}$O (right-panel) as a function 
$r_1=r_2=r$ and the opening angle between the valence neutrons $\theta_{12}$ for settings mentioned in caption of Table~\ref{t1} .}
\label{DenS}
\end{figure}

\vspace{-0.3cm}
The 3b-model with two non-interacting particles in the above single-particle levels of $^{21}$C and $^{25}$O produces
different parity states, when two neutrons are placed in different unbound orbits. The seven configurations, $(s_{1/2})^2$, $(p_{1/2})^2$, $(p_{3/2})^2$, $(d_{3/2})^2$,
$(d_{5/2})^2$, $(f_{5/2})^2$ and $(f_{7/2})^2$ couple to $J^\pi=0^+$ for $^{22}$C and three configurations $(d_{3/2})^2$, $(p_{3/2})^2$ and $(f_{7/2})^2$
couple to $J^\pi=0^+$ for $^{26}$O.
In the 3b-calculations, along with the core-$n$ potential the other important ingredient is the $n$-$n$ interaction.
An attractive contact-delta pairing interaction is used, 
{$g\delta(\vec r_1 - \vec r_2)$} for simplicity, with the only adjustable parameter being $g$. 
We also use the DD contact-delta pairing interaction to explore the  role of different pairing interactions. 
In the DD contact-delta pairing interaction (defined by Eq.~(8) of Ref.~\cite{JS2}), the strength of the DI part is given
as $v_0=2\pi^2\frac{\hbar^2}{m_N}\,\frac{2a_{nn}}{\pi-2k_ca_{nn}}$,
where $a_{nn}$ is the scattering length for the free neutron-neutron scattering
and $k_c$ is related to the cutoff energy, ${{e}_{c}}$, as 
$k_c=\sqrt{\frac{m_Ne_{c}}{\hbar^2}}$.
We use $a_{nn}=−15$\,fm and $e_{c}=30$\,MeV \cite{HAG}, which leads to $v_0=857.2$\,MeV\,fm$^3$. 
For the parameters of the DD part, we determine them so as to fix the ground-state energy of
$^{22}$C and $^{26}$O, $E=-0.140$ MeV \cite{GAUD12} and $0.018$ MeV \cite{KON} respectively. The values of the parameters that we
employ are $R_\rho=1.25\times A_c^{\frac{1}{3}}$ ($A_c=20, 24$), $a=0.65$\,fm. and $v_\rho=591.55$ and $1058.70$\,MeV\,fm$^3$ for $^{22}$C and $^{26}$O respectively. 
We found that configuration mixing in the ground state of $^{22}$C 
does not change much with the choice of $n-n$ interaction. We present the numbers for our potential Set~$1$ in Table~\ref{t1} which are consistent with results of Ref.~\cite{HOR06}, 
and the same behavior is observed for the other sets \cite{JS2}. Whereas for the $^{26}$O case our results report different magnitude of configuration mixing for  the 
different interactions and the results corresponding to the DD part that are consistent with the results of Ref.~\cite{HAG}, where they have also used the DD interaction. 
One possible reason for this difference is that $^{26}$O is unbound by $18$ keV whereas $^{22}$C is bound by $0.140$ MeV. 
In order to reach a final conclusion we need more analysis, which will be reported elsewhere. 
The two particle density of $^{22}$C and $^{26}$O as a function of two radial coordinates, $r_1$ and $r_2$, 
for valence neutrons, and the angle between them, $\theta_{12}$ in the LS-coupling scheme 
is calculated by following Refs.~\cite{JS1,HAG}. The distribution at smaller and larger $\theta_{12}$ are 
referred to as \textquotedblleft di-neutron\textquotedblright 
and \textquotedblleft cigar-like\textquotedblright configurations, respectively. 
One can see in Fig.~\ref{DenS} that the two-particle density is well concentrated around $\theta_{12}\leq90^{\circ}$, which is the clear indication of 
the di-neutron correlation and the di-neutron component has a relatively higher density in comparison to the small cigar-like component for both $^{22}$C and $^{26}$O. 
The reflection of dominance of $s$-component in ground state of $^{22}$C can be seen in left panel of Fig.~\ref{DenS} showing extended di-neutron component 
in comparison to $^{26}$O (in right panel of Fig.~\ref{DenS}), which has sharper dineutron component due to  the mixing of $l>0$ components in its ground-state.
\begin{table}[]
\centering
\vspace*{-0.35cm}
\caption{Components of the ground state ($0^+$) of $^{22}$C and $^{26}$O, with model parameter energy cut $E_{cut}$. For $^{22}$C we have used 
the potential Set~$1$ of Table~1 of \cite{JS2} for the shallow case with the ground-state energy, $-0.140$\,MeV and for $^{26}$O we have used the potential tabulated in Table~\ref{T2}. 
In last column of the table, the comparison has been made with the Ref.~\cite{HOR06} for $^{22}$C and Ref.~\cite{HAG} for $^{26}$O. }
\label{t1}
\begin{tabular}{cccccc}
\hline\hline
 \multicolumn{1}{c}{System}&\multicolumn{1}{c}{$E_{cut}$ (MeV)}& \multicolumn{1}{c}{$lj$}        & \multicolumn{1}{c}{$DI$} &\multicolumn{1}{c}{$DD$}&\multicolumn{1}{c}{Reference}     \\
 \multicolumn{1}{c}{}&\multicolumn{1}{c}{}         &\multicolumn{1}{c}{}   & \multicolumn{2}{c}{\textbf{Present work}}&\multicolumn{1}{c}{}\\    \hline
\multicolumn{1}{c}{}&\multicolumn{1}{c}{}&\multicolumn{1}{c}{$(s_{1/2})^2$}&\multicolumn{1}{c}{$0.923$}&\multicolumn{1}{c}{$0.899$}  &   \multicolumn{1}{c}{$0.915$}\\
\multicolumn{1}{c}{}&\multicolumn{1}{c}{}&\multicolumn{1}{c}{$(p_{1/2})^2$}&\multicolumn{1}{c}{$0.004$}&\multicolumn{1}{c}{$0.008$}&      \multicolumn{1}{c}{$0.009$}\\
\multicolumn{1}{c}{}&\multicolumn{1}{c}{}&\multicolumn{1}{c}{$(p_{3/2})^2$}&\multicolumn{1}{c}{$0.022$}&\multicolumn{1}{c}{$0.029$}&      \multicolumn{1}{c}{$0.024$}\\
\multicolumn{1}{c}{$^{22}$C}&\multicolumn{1}{c}{5}&\multicolumn{1}{c}{$(d_{3/2})^2$}&\multicolumn{1}{c}{$0.045$}&\multicolumn{1}{c}{$0.047$}  &   \multicolumn{1}{c}{$0.033$}\\
\multicolumn{1}{c}{}&\multicolumn{1}{c}{}&\multicolumn{1}{c}{$(d_{5/2})^2$}&\multicolumn{1}{c}{$0.001$}&\multicolumn{1}{c}{$0.005$}&      \multicolumn{1}{c}{$0.003$}\\
\multicolumn{1}{c}{}&\multicolumn{1}{c}{}&\multicolumn{1}{c}{$(f_{5/2})^2$}&\multicolumn{1}{c}{$0.0003$}&\multicolumn{1}{c}{$0.001$}&      \multicolumn{1}{c}{$0.033$}\\
\multicolumn{1}{c}{}&\multicolumn{1}{c}{}&\multicolumn{1}{c}{$(f_{7/2})^2$}&\multicolumn{1}{c}{$0.003$}&\multicolumn{1}{c}{$0.007$}&      \multicolumn{1}{c}{$0.007$}\\\hline\hline
\multicolumn{1}{c}{}&\multicolumn{1}{c}{}&\multicolumn{1}{c}{$(d_{3/2})^2$} &  \multicolumn{1}{c}{0.798} &    \multicolumn{1}{c}{$0.643$} &    \multicolumn{1}{c}{$0.661$}\\
\multicolumn{1}{c}{$^{26}$O}&\multicolumn{1}{c}{10}&\multicolumn{1}{c}{$(p_{3/2})^2$}&  \multicolumn{1}{c}{$0.024$}&     \multicolumn{1}{c}{$0.088$}&     \multicolumn{1}{c}{$0.105$}\\
\multicolumn{1}{c}{}&\multicolumn{1}{c}{}&\multicolumn{1}{c}{$(f_{7/2})^2$} &  \multicolumn{1}{c}{$0.178$}&      \multicolumn{1}{c}{$0.268$}&     \multicolumn{1}{c}{$0.183$}\\\hline\hline
\end{tabular}
\end{table}
\section{Conclusions}
In the present study we present the emergence of bound 2$n$-halo ground state of $^{22}$C from the coupling of seven unbound $spdf$-waves in the
continuum of $^{21}$C and 2$n$-unbound ground state of $^{26}$O from the coupling of three unbound $pdf$-waves in the
continuum of $^{25}$O due to presence of pairing interaction. Contribution of different configurations has been presented along with the 2$n$-correlations. 
More investigations are needed to comment on the role of different pairing interactions and to explain in detail the role of each wavefunction component.

\small{\textbf{Acknowledgements}:
\small J. Singh gratefully acknowledged the financial support from Nuclear Reaction Data Centre (JCPRG), Hokkaido University, Sapporo. 
This work was in part supported by JSPS KAKENHI Grant Numbers 18K03635, 18H04569 and 19H05140, and the collaborative research program 2019,
information initiative center, Hokkaido University.}
\end{document}